\documentclass[]{elsarticle}
\usepackage{amsmath}
\usepackage{setspace}
 
\def\R{\mathcal{R}} 
\def\Q{\mathcal{Q}}
\def\c{{\rm cl}} 
    \newenvironment{myindentpar}[1]%
     {\begin{list}{}%
             {\setlength{\leftmargin}{#1}}%
             \item[]%
     }
     {\end{list}}
\newtheorem{lemma}{Lemma}[section]
\newtheorem{theorem}{Theorem}
\newtheorem{proposition}[lemma]{Proposition}

\title{Minimal autocatalytic networks}

\author[uc]{Mike Steel\corref{cor}}
\ead{mike.steel@canterbury.ac.nz}

\author[smar]{Wim Hordijk}
\ead{wim@SmartAnalytiX.com}

\author[uc]{Joshua Smith}
\ead{joshiansmith@gmail.com}

\cortext[cor]{Corresponding author. Email: {\tt mike.steel@canterbury.ac.nz}, phone: +6421329705.}
\address[uc]{Biomathematics Research Centre, University of Canterbury, Private Bag 4800, Christchurch, New Zealand}
\address[smar]{SmartAnalytiX.com, Lausanne, Switzerland}

\begin{document}

\begin{abstract}
Self-sustaining autocatalytic chemical networks represent a  necessary, though not sufficient condition for the emergence of early living systems. These networks have been  formalised and investigated within
the framework of RAF theory, which has led to a number of insights and results concerning the likelihood of such networks forming. In this paper, we extend this analysis
by focussing on how {\em small} autocatalytic networks are likely to be when they first emerge.   First we show that simulations are unlikely to settle this question, by establishing that the problem of finding a  smallest RAF within a catalytic reaction system is NP-hard.  However, irreducible RAFs (irrRAFs) can be constructed in polynomial time, and we show it is possible to determine in polynomial time whether a bounded size  set of these irrRAFs contain the smallest RAFs within a system.
Moreover, we derive rigorous bounds on the sizes of small RAFs and use simulations to sample irrRAFs under the binary polymer model. 
 We then apply mathematical arguments to prove a new result suggested by those simulations:
at the transition catalysis level at which RAFs first form in this model,  small RAFs are unlikely to be present.    We also investigate further the relationship between RAFs and another formal approach to self-sustaining and closed
chemical networks, namely chemical organisation theory (COT).

\end{abstract}


\maketitle

    \begin{myindentpar}{1cm}
    ``Individual chemical reactions in living beings are strictly coordinated and proceed in a certain sequence, which as a whole forms a network of biological metabolism directed toward the perpetual self-preservation, growth, and self-reproduction of the entire system under the given environmental conditions'' Oparin (1965) \cite{opa}
\end{myindentpar}

\section{Introduction}

 A chemical reaction system that is self-sustaining and collectively autocatalytic is believed to represent
an important step in the emergence of early life \cite{dys, eig, kau1, kau2}.  These systems are defined by two properties: (i) each molecule can be built up from a small subset of pre-existing `food' molecules by some reaction in the system, and (ii) each reaction is catalysed by some product of another reaction (or an element of the food set).  Moreover, recent experimental work has demonstrated at least the possibility (and viability) of such sets \cite{ash, hay, lee, sie, tar, vai}.  It is also of interest to develop a mathematical framework that allows us to study the entire universe of possible self-sustaining autocatalytic sets, so that general results can be established, and predictions made.  Here, we further explore one approach (`RAF theory') which has provided a tractable and incisive tool for addressing computational and stochastic questions. 

RAF theory grew out of two strands: Stuart Kauffman's pioneering work on random autocatalytic networks from the 1970s and 1980s \cite{kau1, kau2, kau3}, and analysis of the first emergence of cycles in random directed graphs by Bollobas and Rasmussen \cite{bol}. Both of these earlier studies were explicitly motivated by origin-of-life considerations. The approach is related to, but different from chemical organisation theory (COT) \cite{con, dit} and other formal approaches of a similar flavour, which  include Petri nets \cite{sha}, Rosen's (M; R) systems \cite{jar,let}, and Eigen and Schuster's hypercycle theory \cite{eig}.
 
 In earlier work \cite{hor} -- \cite{hor4}, \cite{mos, ste} we have established a series of results concerning the structure, discovery and probability of the formation of RAF sets in a variety of catalytic reaction systems. When such a system contains a self-sustaining autocatalytic set (an `RAF', defined below), this set can often be broken down into smaller RAFs until we arrive at the smallest `building block' RAFs that cannot be broken down any further ({\em c.f.} \cite{vas}).  In this paper, we investigate the structure of these irreducible RAFs, and bounds on the size of the smallest RAFs within a catalytic reaction system.  
 
 Along the way, we derive some new facets of RAF theory,  exploring further its relationship to COT, and the
 related weaker notions of pseudo-RAFs and co-RAFs, which can be co-opted by a RAF to form a larger RAF system.  While it is easy to determine whether a chemical reaction system contains an RAF (in which case there is a unique largest one \cite{hor}), we prove that finding a smallest RAF is an NP-hard problem. Nevertheless, the structure of the smallest (`irreducible') RAFs allows us to present efficient algorithms to  find lower bounds on their size, and  to determine whether a given collection contains the smallest RAF in the system.
 
 We begin by recalling some definitions before proceeding to the combinatorial and algorithmic aspects of RAFs. We then apply mathematical arguments and simulations to study the size and distribution of irreducible RAFs in Kauffman's  random binary polymer model \cite{kau3}, and show that at a level of catalysis at which RAFs first form, small RAFs are highly unlikely. We end with a short discussion. 

\section{Definitions}

To formalize the notion of a chemical reaction system (CRS), the following basic notation and definitions are useful:
\begin{itemize}
\item Let $X=\{x_1, x_2, x_3 ,\ldots\}$ be a set of \textit{molecule types}:  each element $x_i$ represents a different type of molecule.
\item Let $F \subset X$ be a \textit{food set}, containing molecule types that are assumed to be freely available in the environment.
\item Let $r = a_1+a_2+\ldots+a_n \rightarrow b_1+b_2+\ldots+b_m$ be a chemical \textit{reaction}, transforming a set of $n$ \textit{reactants} (molecule types $a_1,a_2,\ldots,a_n$) into a set of $m$ \textit{products} (molecule types $b_1,b_2,\ldots,b_m$). In principle there is no restriction on the number of reactants or products, although in the specific model we use (see below) $n$ and $m$ are at most two.
\item Let $\R=\{r_1,r_2,\ldots,r_k\}$ be a set of (chemically possible) reactions.
\item Let $\rho(r)$ and $\pi(r)$ denote, respectively, the set of all reactants of $r$ and the set of all products of $r$, and for any subset $\R'$ of $\R$, let $\rho(\R') = \bigcup_{r \in \R'}\rho(r)$ and $\pi(\R') = \bigcup_{r \in \R'}\pi(r)$.
\item Let $C \subseteq \{(x,r)|x \in X, r \in \R\}$ be a \textit{catalysis set}, i.e., if the molecule-reaction pair $(x,r) \in C$ then molecule type $x$ catalyses reaction $r$.
\end{itemize}

A \textit{chemical reaction system} (or, equivalently, a \textit{catalytic reaction system}; CRS) is now defined as a tuple $\Q=\{X,\R,C\}$ consisting of a set of molecule types, a set of (possible, or allowed) reactions, and a catalysis set. Based on \cite{Bonchev:94}, we can visualise a CRS as a \textit{reaction graph} with two types of vertices (molecules and reactions) and two types of directed edges (from molecules to reactions and vice versa, and from catalysts to the reactions they catalyse).

\subsection{RAF sets}
Informally, a subset of reactions  $\R'$ is an RAF (reflexively-autocatalytic and $F$-generated) set  if  it satisfies the following property:

    \begin{myindentpar}{1cm}
Every reactant of every reaction in $\R'$ can be built up by starting from $F$ and using just reactions in $\R'$, and so that all reactions are eventually catalysed by at least one molecule that is either a product of some reaction in $\R'$ or is an element of $F$.
    \end{myindentpar}
To define an autocatalytic set more formally, we first need to define the notion of ``closure''. Informally, the closure of a set of molecule types relative to a set of reactions, is the initial set of molecule types together with all the molecule types that can be created from it by repeated application of reactions from the given set of reactions. More formally, given a CRS $\Q=\{X,\R,C\}$, the \textit{closure} $\c_{\R'}(X')$ of $X' \subseteq X$ relative to $\R' \subseteq \R$ is the (unique) minimal set $W \subseteq X$ that contains $X'$ and satisfies the condition that, for each reaction $r=A \rightarrow B \in \R'$ (with $A$ being a set of reactants and $B$ a set of products),
$A \subseteq W \Longrightarrow B \subseteq W.$ 
Notice that when $\R' = \emptyset$ the set  $\c_{\R'}(X')$ is still defined, and it equals $X'$.

Our mathematical definition of RAF sets is now as follows (note that this is the definition from \cite{hor2}, which is slightly modified from the original definition in \cite{hor}). Given a CRS $\Q=\{X,\R,C\}$ and a food set $F \subset X$, a non-empty subset $\R' \subseteq \R$ is said to be:
\begin{itemize}
\item {\it Reflexively autocatalytic} if, for all reactions $r \in \R'$, there is at least one molecule type $x \in \c_{\R'}(F)$ such that $(x,r) \in C$;
\item {\it $F$-generated} if $\rho(\R')  \subseteq \c_{\R'}(F)$;
\item {\it Reflexively autocatalytic and $F$-generated} (RAF) for $(\Q,F)$ if $\R'$ is both reflexively autocatalytic and $F$-generated.
\end{itemize}

\begin{figure*}
\center
\resizebox{8cm}{!}{
\includegraphics{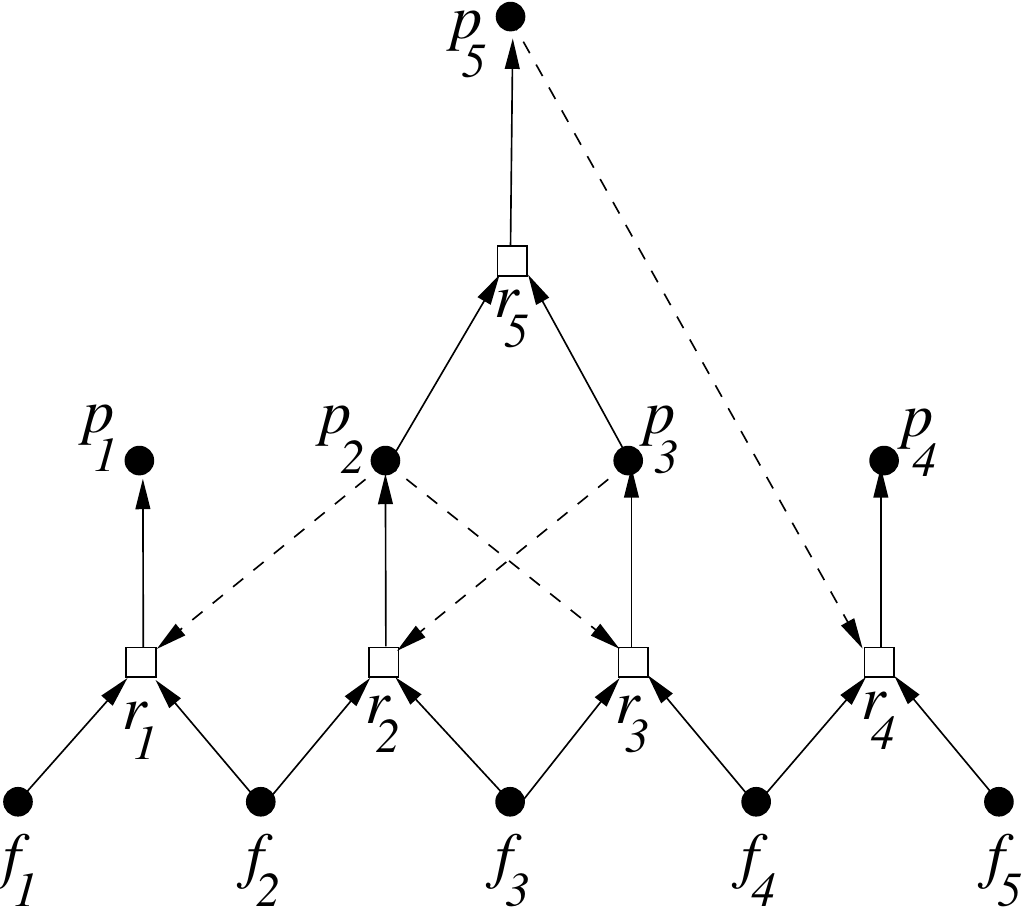}
}
\caption{A CRS for which the maxRAF consists of the set of three reactions $\{r_1, r_2, r_3\}$.  The only other RAF present is the irrRAF $\{r_2, r_3\}$.  The singleton 
reaction $\{r_1\}$ is not an RAF (but it forms a co-RAF, defined later).}
 \label{figure-u}
\end{figure*}

Because the union of RAFs for $(\Q,F)$ is also an RAF for $(\Q,F)$ it follows that any  CRS that contains an RAF has a unique maximal RAF called the  `{\rm max}RAF';
any other RAF is called a `subRAF' of this maximal RAF.  We say that an RAF is an  {\em irreducible RAF} (or, more briefly, an `irrRAF')  if no proper subset is also an RAF. In contrast to the uniqueness of the maximal RAF, there
may be many  (indeed exponentially many) irrRAFs  \cite{hor3}.

\section{Characterising $F$-generated sets}

We have already defined the concept of  being$F$-generated,  however, it will be useful to explore this further for the following reasons:
\begin{itemize}
\item to better understand the distinction between RAFs and `pseudo-RAFs' (defined shortly);
\item to explain the link between $F$-generated sets and `organisations' in chemical organisation theory;
\item to provide a characterisation that we will require later in the proof of our main stochastic theorem (Theorem~\ref{nosmall}).
\end{itemize}

Given a CRS $\Q= (X, \R, C)$ and a food set $F$,  the closure set ${\rm cl}_{\R'}(F)$ has two further equivalent descriptions. Firstly, it is 
the intersection of all subsets of $X$ that contain $F$ and that are closed relative to $\R'$.  It  also has an explicit constructive definition as follows:  ${\rm cl}_{\R'}(F)$ is the final set 
$W_K$ in the  sequence of nested sets $F = W_0 \subseteq W_1 \subseteq \cdots \subseteq W_K$  where
$W_{i+1}$ is equal to the union of $W_i$ and  the set of products of reactions in $\R'$ whose reactants lie in $W_i$, and where
$K$ is the first value of $i$  for which $W_{i} = W_{i+1}$.

\begin{figure*}[ht] 
\center
\resizebox{12cm}{!}{
\includegraphics{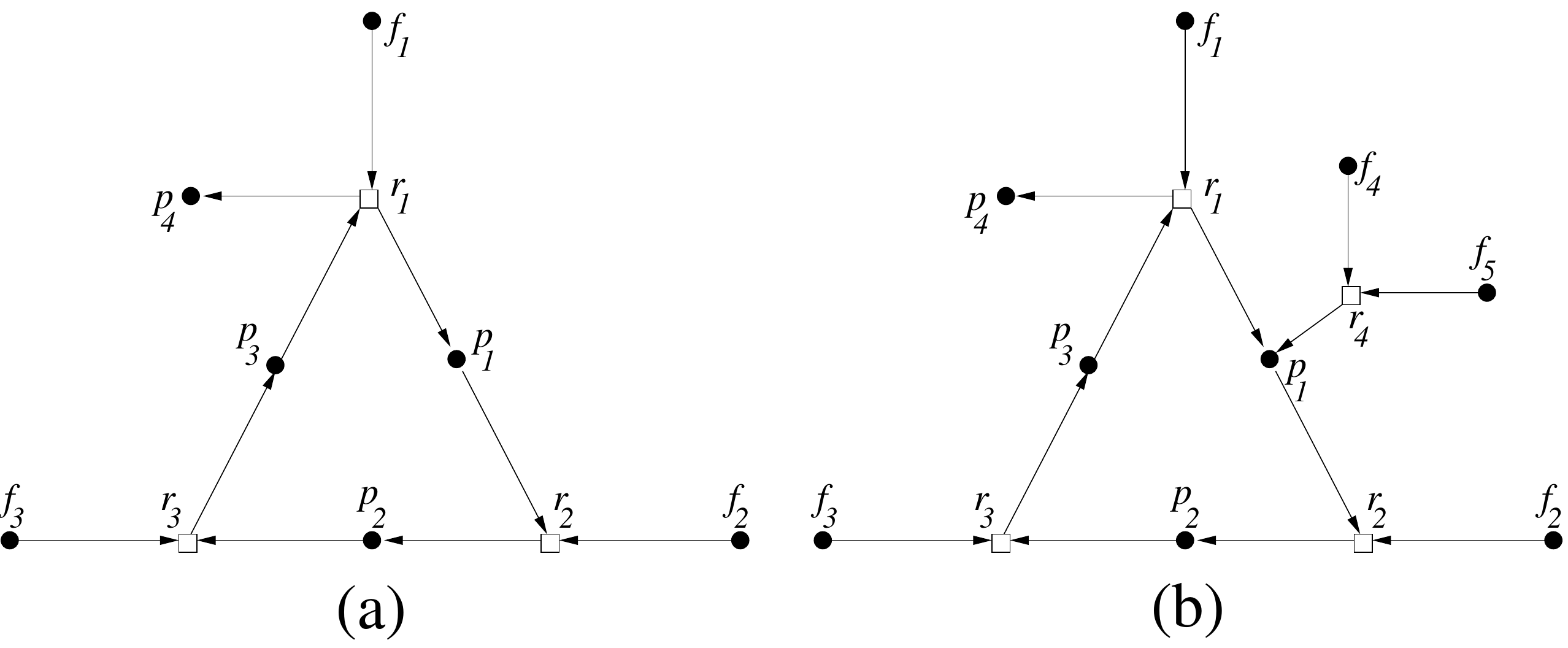}
}
\caption{(a) The set $\R' =\{r_1, r_2, r_3\}$ of reactions is not $F$-generated, for $F= \{f_1, f_2, f_3\}$ and $X=F \cup \{p_1, p_2, p_3, p_4\}$.  (b) The expanded reaction set  $\R =\{r_1, r_2, r_3, r_4\}$ is $F$-generated, for  $F= \{f_1, f_2, f_3, f_4, f_5\}$.  The (unique) ordering that satisfies the conditions of Lemma~\ref{lemequiv2}(iii) (or (iv)) is  $r_4, r_2, r_3, r_1$.}
 \label{figure_x}
\end{figure*}

With this in hand, we now examine  the definition of $F$-generated sets of reactions more closely.
Recall from the earlier definitions that a  subset of reactions $\R'$ is $F$-generated  provided that every reactant of every reaction in $\R'$ lies in ${\rm cl}_{\R'}(F)$.   
Note that saying $\R'$ is $F$-generated implies but is {\bf strictly stronger} than the condition that the reactant of each reaction in $\R'$ is either a molecule  in  $F$ or it  is a product of another reaction in $\R'$.  $F$-generated is also strictly stronger than requiring that the molecules of $X$ that are `used up' in maintaining the reactions in $\R'$ is precisely $F$. An example that demonstrates both these strict containments is provided in Fig.~\ref{figure_x} for the set $\R' = \{r_1, r_2, r_3\}$, which is not $F$-generated (since $\c_{\R'}(F) = F)$.

 We now provide precise characterizations of when a set of reactions is $F$-generated.   

\begin{lemma}
\label{lemequiv2}
Given a CRS $\Q= (X, \R, C)$, a  food set $F$ and a non-empty subset $\R'$ of $\R$, the following are equivalent:
 \begin{itemize}
\item[(i)]
$\R'$ is $F$-generated.
\item[(ii)] 
$\c_{\R'}(F) = F \cup \pi(\R')$.
\item[(iii)]
$\R'$ has a linear ordering $r_1,\ldots, r_k$ so that the reactants of $r_1$ are molecules in $F$, and for each  $i \in \{2,\ldots, k\}$ the reactants of $r_{i}$ are contained in $\c_{\{r_1, \ldots, r_{i-1}\}}(F).$
\item[(iv)]
$\R'$ has a linear ordering $r_1,\ldots, r_k$ so that 
 for each $i \in \{1,\ldots, k\}$ each reactant of $r_{i}$ is either an element of $F$ or is a product of some reaction $r_j$ where $1\leq j < i$.
\end{itemize}
 \end{lemma}

 {\em Proof:}   The equivalence $(i) \Leftrightarrow (ii)$ is from \cite{hor} (Lemma 4.3) and the equivalence  $(iii) \Leftrightarrow (iv)$ is easily verified, as the ordering of $\R$
 that applies for either part, also works for the other (from the definitions).   Thus, to establish this four-way equivalence, it suffices to show that $(i) \Rightarrow  (iii)$, and $(iii) \Rightarrow (i)$.

To establish $(i) \Rightarrow (iii)$, suppose that  $\R'$ is $F$-generated.  We construct an ordering satisfying $(iii)$ as follows:  Let $\R_0$ denote the reactions in $\R'$ that have their  reactants in $F$, and for $i>0,$ let $\R_i$ denote the reactions in $\R'$ that have their reactants in the set $W_i - W_{i-1}$, where $W_i, i \geq 0$ is the sequence of nested
 sets described in the preamble to this lemma.  Then take any ordering on $\R'$ for which the reactions in $\R_i$ all come before $\R_{i+1}$ for $i=0,\ldots, K-1$. This ordering satisfies the property described in part $(iii)$.

To establish $(iii) \Rightarrow (i)$, we only need to observe that $\c_{\{r_1, \ldots, r_{i-1}\}}(F)  \subseteq \c_{\R'}(F)$ for all $i>1$, so if $\rho(r_i)$ is a subset of the first set, it is necessarily a subset of the second set.  This completes the proof of Lemma~\ref{lemequiv2}.

 \bigskip
 
We now point out a consequence of this lemma that sheds some light on why the subset $\R'$ in Fig.~\ref{figure_x} fails to be $F$-generated. 
Given a CRS $\Q= (X, \R, C)$, a food set $F$ and a subset  $\R'$ of $\R$, consider the directed graph $G(\R')$ that has vertex set $\R'$ and an arc from $r$ to $r'$ precisely if there is a reactant $x$ of $r'$ that is a product of $r$ and, in addition, if $x \not\in {\rm cl}_{\R'-\{r\}}(F)$. This last condition states that molecule $x$ cannot be built up 
from $F$ using only the reactions in $\R'$ that do not include $r$.   Note  that a vertex of  $G(\R')$ is permitted to have  a loop  (i.e.  an arc from a reaction to itself).  
As an example of this graph, for the reactions shown in Fig.~\ref{figure_x}(a),
$G(\R')$ is a directed three-cycle, while in part (b) of that figure, $G(\R')$ has no directed cycle.

 \begin{theorem}
 \label{corol}
 Given a CRS $\Q= (X, \R, C)$,  a food set $F$, a non-empty subset $\R'$ of $\R$ is $F$-generated if and only if the following two conditions hold:
\begin{itemize}
\item[(a)]
every reactant of a reaction in $\R'$ is either an element of $F$ or is a product of some reaction in $\R'$; and 
\item[(b)]
the graph $G(\R')$ has no directed cycle (including loops).
\end{itemize}
\end{theorem}

{\em Proof:}  Suppose that $\R'$ is $F$-generated. Then  condition (a)  in the theorem follows by part (ii) of Lemma~\ref{lemequiv2}; moreover, there exists an ordering $r_1, \ldots, r_k$  of $\R'$ that satisfies the condition described in part (iii) of that lemma.  Now,  if  $(r_i, r_j)$ is an arc in $G(\R')$, we must have $i<j$, since otherwise, if $i \geq j$,   part (iii) of  
 Lemma~\ref{lemequiv2}   gives:
$$\rho(r_j) \subseteq \c_{\{r_1, \ldots, r_{j-1}\}} (F) \subseteq \c_{\R'-\{r_i\}}(F),$$
and the containment  $\rho(r_j) \subseteq \c_{\R'-\{r_i\}}(F)$ would preclude the arc $(r_i, r_j)$ from  $G(\R')$.  So, if $G(\R')$ had a directed cycle $(i_1, i_2), (i_2, i_3), \ldots, (i_r, i_1)$,  we would have: $i_1<i_2 < \ldots < i_1$, a contradiction.  Thus if $\R'$ is $F$-generated, condition (b) in the theorem also holds.

Conversely, suppose that $\R'$ satisfies conditions (a) and (b).  We first show that there exists a reaction $r^* \in \R'$ that has all its reactants in $F$, i.e. $\rho(r^*) \subseteq F$.
Suppose to the contrary that this were not the case (we will show this contradicts condition (b)). Then for every reaction $r$ in $\R'$, we can select a molecule $x(r) \not\in F$ that is a reactant of $r$.  Moreover, by property (a) and the condition that
$x(r) \not\in F$ it follows that $x=x(r)$ is the product of some other reaction, which we will write as $r'(x)$.  Thus, starting with any given reaction, $r_0$, consider the alternating sequence of molecules and reactions $(x_i, r_i), i \geq 0$ that we generate from $r_0$ by setting $x_{i} = x(r_{i})$ and $r_{i+1} = r'(x_{i})$. 
Since $\R'$ is finite, this sequence must have $r_k=r_l$ for some $0 \leq k<l$.
Moreover, we cannot have $x_i \not\in \c_{\R'-\{r_{i+1}\}}(F)$ for all  $i \in [k, l-1]$; otherwise, in the graph $G(\R')$,  there would be an arc from $r_{i+1}$ to $r_i$ for all 
$i \in [k,l-1]$ and so we would obtain a directed cycle in $G(\R')$, and by part (b), no such cycle exists. This contradiction ensures there exists some molecule $x_i \not\in F$ for $i \in [k,l-1]$ for which  $x_i \in \c_{\R'-\{r_{i+1}\}}(F)$. However, if the closure of $F$ under  {\em any} set of reactions contains a molecule outside of $F$, then some reaction in the collection must have all its reactants in $F$ (by Lemma~\ref{lemequiv2}).  This justifies our claim that there is a reaction $r^* \in \R'$ with $\rho(r^*) \subseteq F$.

We now use induction on $|\R'|$ to establish that conditions (a) and (b) imply that $\R'$ is $F$-generated.  For $|\R'|=1$ and the non-existence of a loop from this reaction to itself (by (b)), we see that $\R'$ is $F$-generated.  Therefore suppose that the implication holds  for any $|\R'| <n$ satisfying (a) and (b), and that we have $|\R'|=n$.  Now, consider $\R'' = \R'-\{r^*\}$ and $F' = F \cup \pi(r^*)$, where $r^*$ is the reaction in $\R'$ with $\rho(r^*) \subseteq F$.
Notice that $\R''$ satisfies property (a). Moreover, we claim that property (b) also holds for
$\R''$ since if $(r, r')$ is an arc of $G(\R'')$ then it is also an arc of $G(\R')$. To verify this, observe that if $(r, r')$ is an arc of $G(\R'')$ then there exists a reactant $x$ of $r'$ that is a product of
$r$ and for which $x \not\in \c_{\R''- \{r\}}(F')$.  However:  $$\c_{\R''- \{r\}}(F')= \c_{\R' - \{r,  r^*\}}(F') = \c_{\R'-\{r\}}(F),$$ and so $x \not\in  \c_{\R'-\{r\}}(F)$, which implies that $(r,r')$ is indeed
an arc of $G(\R')$.  Consequently,  the arcs of $G(\R'')$ are a subset of the set of arcs of $G(\R')$ that do not contain $r^*$
and so $G(\R'')$ cannot contain a directed cycle (or else $G(\R')$ would).   

Thus, since $\R''$ satisfies properties (a) and (b), it follows (by the induction hypothesis) that $\R''$ is $F'$-generated, and this
implies that $\R'$ is  $F$-generated. This completes the proof of the converse result.

  \bigskip

\section{Relationship with chemical organisation theory (COT)}

Chemical organisation theory (COT) \cite{dit} provides another way to study chemical reaction systems, and the concept of a (chemical) {\em organisation} shares two 
key properties with RAFs: closure and self-maintenance (for precise definitions, see \cite{dit}, and for recent relevant results, see \cite{con} and \cite{kre}).   However, the latter concept (`self maintenance') is defined somewhat differently: while RAFs require the property of being $F$-generated, an organisation is defined as self-sustaining under chemical dynamics, as encoded by the stoichiometric matrix.  More precisely, if ${\bf S}$ is the stoichiometric matrix for the system, with its rows indexed by molecules  and its columns by reactions, then self-maintenance requires a column vector ${\bf v}$ with strictly positive coordinates for which:
\begin{equation}
\label{sm}
{\bf Sv} \geq {\bf 0}
\end{equation}
 In words, this is the condition that the reactions can proceed at positive rates,  so that the net rate of production of each molecule in the system  is not less than the rate at which it is used up (otherwise such a molecule would disappear from the system).  This is a weaker requirement than being $F$-generated, since self-maintenance requires only that the system be self-sustaining once it exists, but does not address the question of whether the system could form in the first place from a set of molecules in $F$;   we describe an example to illustrate this shortly.  

A second difference is that organisations allow but do not explicitly require
reactions to be catalysed, though an extension to allow this has been discussed recently in \cite{con}. Note that RAFs easily extend to allow some reactions not to be catalysed by introducing a putative new element of $F$ to act as a catalyst for any reactions that otherwise do not require catalysis.

A third important difference is algorithmic and we will discuss this shortly (a further minor difference is that organisations are subsets of molecules, while an RAF is a subset of reactions and molecules).   The following lemma shows that there is a close but not identical relationship  between $F$-generated sets and organisations; part (i) was discovered by \cite{con}.

\begin{lemma}
\mbox{ } 
Given a CRS $\Q=(X, \R, C)$ and food set $F$, consider the set $\R_F:=\{\emptyset \rightarrow f: f \in F\}$ of reactions that formally generate $F$ without using other molecules in $X$.
\begin{itemize}
\item[(i)] If $\R'$ is $F$-generated then the set of molecules  ${\rm cl}_{\R'}(F)$ forms an organisation, for the reactions $\R' \cup \R_F$.
\item[(ii)]
It is possible for a set $M$ of molecules to form an organisation for a set of reactions $\R' \cup \R_F$ but for $\R'$ to fail to be $F$-generated.
\end{itemize}
\end{lemma}

Part (i) of the lemma was established in \cite{con} (Corollary 1). 
Here, we show how it also follows as a consequence of Lemma~\ref{lemequiv2}. 
Firstly, if $\R'$ is $F$-generated, then it is closed by the implication (i) $\Rightarrow$ (ii)  in  Lemma~\ref{lemequiv2}. 
Moreover, we may order the reactions in $\R' \cup \R_F$ so that the reactions  in $\R_F$ come first (in any order) and so that the order of the subsequent reactions from $\R'$ is such that the reactants of each reaction are either elements of $F$ or products of earlier reactions -- the existence of such an ordering for $\R'$
is provided by the implication (i) $\Rightarrow$ (iv)  in  Lemma~\ref{lemequiv2}.  Consider the corresponding stoichiometric matrix ${\bf S}$.  Then 
the first non-zero element in each row of ${\bf S}$ is +1.  Now for {\em any}  real matrix  with this last property, there is a strictly positive column  vector ${\bf v}$ for which 
${\bf Sv}>{\bf 0}$, since if ${\bf S}$ has $c$ columns, and if the largest absolute value of any negative entry of ${\bf S}$ is $b$ then we can take ${\bf v}$ to be the strictly positive
vector that has  its $i$-th coordinate given by: $v_{c-i} = (b+1)^i$ for $i=0, \ldots, c-1.$

 Part (ii) is established by considering the example shown in Fig.~\ref{figure_x}(a) with $M=F \cup \{p_1, p_2, p_3, p_4\}$ and $\R' = \{r_1, r_2, r_3\}$. 
 Ordering $M$ as $f_1, f_2, f_3, p_1, p_2,       p_3, p_4$ and $\R' \cup \R_F$ as $\emptyset \rightarrow f_1, \emptyset \rightarrow f_2, \emptyset \rightarrow f_3,
 r_1, r_2, r_3$,  we obtain the following $7 \times 6$ stoichiometric matrix (rows are indexed by molecules; columns, by reactions):
 
 $${\bf S} = 
\left[ \begin{array}{rrrrrrr}
1 & 0 & 0 & -1 & 0  & 0 \\
0 & 1 & 0 & 0 & -1  & 0 \\
0 & 0 & 1 & 0 & 0  & -1\\
0 & 0 & 0 & 1 & -1 & 0 \\
0 & 0 & 0 & 0 & 1  & -1 \\
0 & 0 & 0 & -1 & 0 & 1\\
0 & 0 & 0 & 1 & 0  & 0
 \end{array} \right]
$$

It is now clear that ${\bf S}{\bf v} = [0,0,0,0,0,0,1]^T$ for the strictly positive vector ${\bf v} = [1,1,1,1,1,1]^T$ and therefore the self-maintenance inequality (\ref{sm}) holds.  Since $M$ is closed relative to the six reactions, it follows that $M$ forms an organisation, but $\R'$ fails to be $F$-generated, since ${\rm cl}_{\R'}(F)=F$.  This completes the proof.

\bigskip

A further difference between COT and RAF theory is that determining whether or not a CRS contains a non-empty organisation is an NP-complete problem ({\em c.f.} \cite{cen}, Section 6.2), while determining whether there exists an RAF (necessarily non-empty) within any CRS can be decided by a polynomial time algorithm.   We describe this now.

\subsection{The RAF algorithm and the map $\R' \mapsto s(\R')$}
\label{rafalg}

The usual RAF algorithm (\cite{hor, hor2}) starts with the full set of reactions and
iteratively prunes out reactions until the set stabilises. For completeness, we describe this explicitly now.
Given a CRS $\Q= (X, \R, C)$ and a food set $F$, define the following nested (decreasing) sequence of subsets of reactions $\R_0, \R_1, \ldots, R_K$ as follows:
\begin{itemize}
\item $\R_0 = \R$; and for $i\geq 0$, 
\item $\R_{i+1} = \{r \in \R_i: \mbox{ $r$ has all its reactants and at least one catalyst in } {\rm cl}_{\R_i}(F)\}$;
\item $K$ is the first value $i$ for which $\R_i = \R_{i+1}$.
\end{itemize}
It can be shown that if  $\R_K = \emptyset$ then $\R$ contains no RAF; otherwise, $\R_K$ is the unique maximal RAF contained in $\R$ (for further details, see \cite{hor, hor2}).  Throughout this paper, we will let $s(\R')$ denote the terminal set ($\R_K$) obtained by applying this process to an arbitrary subset $\R'$ of $\R$.

\section{Pseudo-RAFs and co-RAFs}

Note that an RAF $\R'$ for $(\Q, F)$ satisfies the following two properties:

\begin{itemize}
\item[(i)]  Every reaction in $\R'$ is catalysed by the product of another reaction from $\R'$ or by an element of $F$; and
\item[(ii)] Each reactant of every reaction in $\R'$ is either an element in $F$ or a product of another reaction in $\R'$.
\end{itemize}

\begin{figure*}[ht] 
\center
\resizebox{8cm}{!}{
\includegraphics{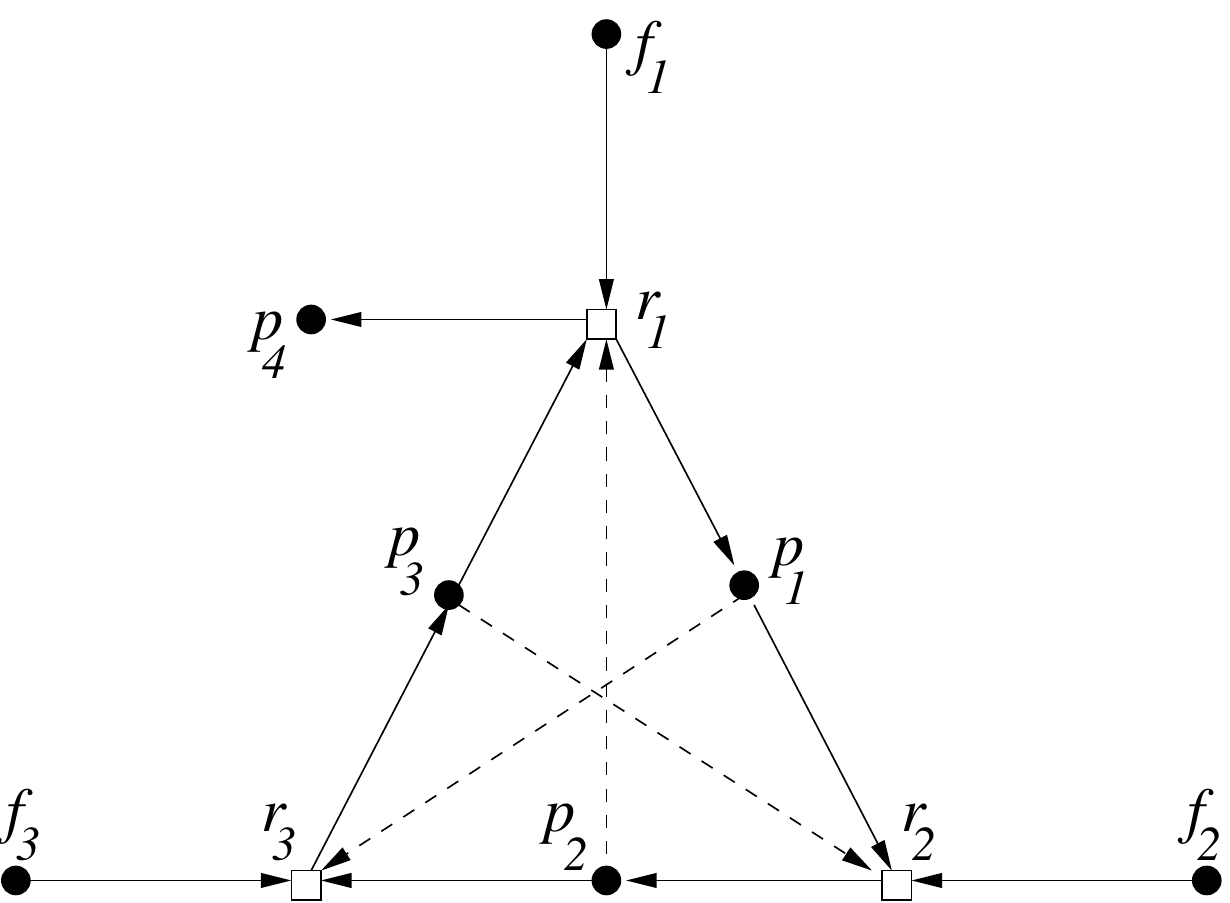}
}
\caption{A pseudo-RAF which fails to be an RAF}
 \label{figure_y}
\end{figure*}

We will call any subset $\R'$ of $\R$ that is non-empty and  that  satisfies properties (i) and (ii) a {\em pseudo-RAF} for $(\Q, F)$.  Not every pseudo-RAF is an RAF, as the example in Fig~\ref{figure_y}) shows.   
However, pseudo-RAFs satisfy some of the properties of RAFs; in particular, the union of two or more pseudo-RAFs for $(\Q,F)$ is a pseudo-RAF for $(\Q,F)$. It follows that any pair $(\Q, F)$ either contains no pseudo-RAF (in which case $(\Q, F)$  contains no RAF either) or $(\Q, F)$  has a unique maximal pseudo-RAF that contains all other pseudo-RAFs of $(\Q,F)$ as well as the unique maximal RAF for  $(\Q,F)$.

An analogous algorithm to the RAF algorithm applies for constructing the maximal pseudo-RAF (when it exists), the only change being that ${\rm cl}_{\R_i}(F)$ is replaced by 
$F \cup \pi(\R_i)$ in the construction of $\R_{i+1}$ from $\R_i$ (where $\pi(\R_i)$ is the set of products of reactions in $\R_i$).

The RAF and pseudo-RAF algorithms have a similar flavour to the  `unit propagation' method of solving the propositional logic problem `HORN-SAT', and in  \cite{hor_ext} we showed that HORN-SAT can be solved by an extension of the RAF algorithm.

\subsection{co-RAFs}

Although a pseudo-RAF cannot become established by itself (since it is not $F$-generated), it can nevertheless become established in the presence of another RAF.  
This property is not unique to pseudo-RAFs, and we formalise and investigate this notion as follows.  

Given a CRS $\Q=(X, \R, C)$ and a food set $F$,  we will say that a subset $\R'$ of $\R$ is a {\em co-RAF} for $(\Q,F)$ if $\R'$ is a non-empty set for which there exists some 
RAF $\R_1$ for $\Q$,  which is disjoint from $\R'$ and whose union with $\R'$, $\R_1  \cup \R'$, forms an RAF for $\Q$.

A simple example of a   co-RAF is the set $\{r_1\}$ in Fig.~\ref{figure-u}.  Informally, a co-RAF is a system that may not have enough structure to form an RAF by itself, but which another (disjoint) RAF can co-opt to form a larger RAF.   Note that a co-RAF may fail to be an RAF because either a reactant or a catalyst (or both) can fail to be in the closure of $F$;  in either case, $\R_1$ can provide the missing $F$-generated reactant or catalyst.

The relationship between an RAF $\R_1$ and an associated co-RAF $\R'$ is similar to the relationship between a `viable core'  and an associated `periphery' in \cite{vas}.
The  requirement that $\R'$ and $\R_1$ are disjoint  in the definition of a co-RAF is not a serious restriction, since if 
$\R' \cup \R_1$ is an RAF for $(\Q,F)$, where $\R'$ overlaps (but is not strictly contained within) an RAF $\R_1$, then $\R'- (\R' \cap \R_1)$ is a co-RAF for $(\Q, F)$.

Determining whether a given subset $\R'$ of $\R$ is a co-RAF for $(\Q, F)$ can be solved in polynomial time by virtue of the following result (the equivalence of parts (i) and (ii)). We also give two other alternative descriptions of co-RAFs.  The proofs of these results are presented in the Appendix.

\begin{proposition}
\label{proprop}
\mbox{ } 
 Given a CRS $Q = (X,\R, C)$ and a food set $F$, let  $\R'$ be a non-empty subset of $\R$. The following are equivalent:
\begin{itemize}
\item[(i)]   $\R'$ is a co-RAF for $(\Q, F)$;
\item[(ii)]  $s(\R-\R') \neq \emptyset$ and $\R' \cup s(\R-\R')$ is an RAF for $(\Q, F)$;
\item[(iii)] $\R' = \R_B-\R_A \mbox{  for two RAFs  }  \R_A, \R_B \mbox{ for } (\Q, F), \mbox{ where }  \R_A \subset \R_B;$ 
\item[(iv)]  $\R'$ is an RAF for $(\Q, F')$ where $F'= F \cup \pi(\R_1)$, for some RAF $\R_1$ 
 for $(\Q, F)$ that is disjoint from $\R'$.
 \end{itemize}
\end{proposition}

Note that the equivalence $(i) \Leftrightarrow (iii)$ provides a simple way to generate co-RAFs:  any non-maximal  RAF $\R_A$ for $\Q$ has a co-RAF; simply let $\R_B$ be the maximal RAF, and take $\R' = \R_B- \R_A$.

\bigskip

\section{Minimal RAFs and irrRAFs}

Given a CRS $\Q= (X, \R, C)$ and a food set $F$, we can  find an irrRAF efficiently  (i.e. in polynomial time), but finding a minimal-sized RAF is much harder -- we will show it is an NP-hard problem even to determine this minimal size.
Nevertheless,  it is possible to test whether a given  irrRAF for $(\Q, F)$  is the only irrRAF for $(\Q, F)$  -- and if it is, then  it is necessarily a minimal sized RAF. More generally, if we generate irrRAFs for $(\Q, F)$ (each in polynomial time), and have found only a relatively small number of them (e.g. $<10$ or so) then it is possible to test whether these are the only irrRAFs for $(\Q,F)$ and, if so, the  one(s) of smallest size are the minimal-sized RAFs for $(\Q, F)$.  We now show how this can be solved efficiently (in polynomial time), provided that we bound the number of irrRAFs.

\subsection{Do we have all the irrRAFs?}

Suppose that a CRS $(X, \R, C)$ with a food set $F \subseteq X$, has an RAF.   Let $\R_1, \R_2,\ldots \R_k$ be a collection of distinct  irrRAFs that have been constructed from this RAF (e.g. by our search algorithm). We would like to be  able to determine whether these are {\em all} the irrRAFs for $(X, \R, C, F)$. The following result provides a way to do this for moderate values of $k$.  Recall that for a subset $\R'$ of $\R$, $s(\R')$ is the result of applying the RAF algorithm to $\R'$.

\begin{theorem}
Suppose that a CRS $(X, \R, C)$ with a food set $F$, has an RAF. Then a collection $\R_1, \ldots, \R_k$ of distinct irrRAFs  constitutes the set of all the irrRAFs for $(X, \R, C, F)$ if and only if the following condition holds:

\bigskip

\noindent For all $(r_1, r_2, \ldots, r_k) \in \R_1 \times \R_2 \times \cdots \times \R_k$, we have
$s(\R-\{r_1, r_2, \ldots, r_k\}) = \emptyset.$
\end{theorem}
{\em Proof: }  Suppose first that for some $(r_1, r_2, \ldots, r_k) \in \R_1 \times \R_2 \times \cdots \times \R_k$ we have $s(\R-\{r_1, r_2, \ldots, r_k\}) \neq  \emptyset$.  Then
$s(\R-\{r_1, r_2, \ldots, r_k\})$ is an RAF and so it contains at least one irrRAF, say $\R'$.  Since $$\R' \subseteq s(\R-\{r_1, r_2, \ldots, r_k\}) \subseteq \R-\{r_1, r_2, \ldots, r_k\},$$
$\R'$ cannot equal $\R_i$ for any $i$, since $\R'$ does not contain $r_i$, but $\R_i$ does. Thus, $\R_1, \ldots, \R_k$ does not constitute the
set of all irrRAFs of $(X, \R, C, F)$.

Conversely, suppose that $\R_1, \ldots, \R_k$ is not the 
set of all irrRAFs. Let $\R'$ be any other irrRAF.  Then  $\R_i$ is not strictly contained within $\R'$ for any $i$ because otherwise $\R'$ would not be an irrRAF.  Thus for each $i$, there
exists some reaction $r_i \in \R_i -\R'$ and thus a sequence $(r_1, r_2, \ldots, r_k) \in \R_1 \times \R_2 \times \cdots \times \R_k$.  Now consider
$s(\R-\{r_1, r_2, \ldots, r_k\})$. Since $\R'$ is a subset of $\R-\{r_1, r_2, \ldots, r_k\})$,  it follows that $$\R'=s(\R') \subseteq s(\R-\{r_1, r_2, \ldots, r_k\}),$$ and so $s(\R-\{r_1, r_2, \ldots, r_k\}) \neq \emptyset$.
This completes the proof.

\bigskip

{\bf Remark:} 
For any given value of $k$, determining whether or not we have all the irrRAFs can be solved in polynomial time (in the size of the CRS).  Of course, the exponent in the polynomial involves $k$,
so it would also be interesting to see if this exponential dependency on $k$ can be removed  (and, if not, whether the problem is fixed parameter tractable  in $k$).

\subsection{Finding a smallest RAF is hard}

Given a CRS and a food set $(\Q, F)$, finding a largest RAF can be solved by a polynomial time algorithm. This raises an obvious question: is there an efficient way to find the {\em smallest} RAF for $(\Q,F)$, or at least to calculate its size? A related  question replaces `smallest RAF' with `smallest irrRAF',  but it is clear that any smallest RAF must also be irreducible so the two questions are equivalent.  Consider then the decision problem: 
\bigskip

\indent{{\bf MIN-RAF}

INSTANCE:  A catalytic reactions system and food set $(X, \R, C, F)$, and a 

positive integer $k$.

QUESTION: Does $\R$ contain a subset of size at most $k$ that forms an RAF 

for $(X, \R, C, F)$?}

\begin{theorem}
\label{npthm}
\mbox{ }
\begin{itemize}
\item[(i)]
The decision problem MIN-RAF is NP-complete.
\item[(ii)]
Counting the number of sub-RAFs  (or smallest sub-RAFs) of an arbitrary RAF  is \#P-complete. 
\end{itemize}
\end{theorem}
\bigskip

The  proof of this theorem involves a reduction of MIN-RAF to the graph theory problem VERTEX COVER, by associating with each CRS a graph that has its vertex covers of size
$K$ in one-to-one correspondence with the sub-RAFs of the CRS of size  $K$+constant.  The details of the construction and the full proof of Theorem~\ref{npthm} are provided in the Appendix.

\subsection{Lower bounds on the size of RAFs}

In the light of Theorem~\ref{npthm}, an interesting question is whether we can efficiently compute lower bounds on the size of an RAF.  The first lower bound is easily computed.\begin{lemma}
Consider a catalytic reaction system $\Q$ and a food set $F$.
Let $$\R_0 = \{r \in \R: s(\R - \{r\}) = \emptyset\}.$$
Then every RAF for $(\Q, F)$ has size at least $|\R_0|$.
\end{lemma}
{\em Proof:}  Let $\R'$ be an RAF for $(\Q, F)$.   Suppose that $r \in \R_0$. If $r \in \R-\R'$ then $\R' = s(\R') \subseteq s(\R-\{r\}) = \emptyset$, which is not possible, since an RAF is non-empty, by definition.  Thus, $r\in \R'$. Since this holds for all $r \in \R'$ it follows that $\R_0 \subseteq \R'$, and so $|\R_0| \leq |\R'|.$ This completes the proof.

\bigskip

Part (ii) of the following Lemma provides a further computable lower bound on the smallest RAF,  if we require the  RAF to have the additional property that none of its
reactions are catalysed by a food molecule.  Given a CRS  $\Q$, let $G'(\R')$ be the graph with the vertex set $\R'$ and with an arc
from reaction $r$ to reaction $r'$ precisely if some product of $r$ is a catalyst of $r'$.   Part (i) of the lemma is essentially the `Loop Theorem' of \cite{con} (Theorem 2).

\begin{lemma}
\label{nof}
Consider a catalytic reaction system $\Q=(X, \R, C)$ and a food set $F$.
\begin{itemize}
\item[(i)]
If $\R'$ is an RAF for $(\Q, F)$ and no reaction in $\R'$ is catalysed by any food molecule then $G'(\R')$ contains a directed cycle.  
\item[(ii)]
Provided that $s(\R) \neq \emptyset$ (i.e. $(Q,F)$ has an RAF), the smallest RAF for $(\Q, F)$ for which no reaction is catalysed by a food molecule is at least as large as the length of the shortest directed cycle in $G'(s(\R))$, and this can be computed in polynomial time in the size of $\Q$. 
\end{itemize}
\end{lemma}

{\em Proof:}    {\em Part (i):}   A classic, elementary result ({\em c.f.} \cite{ban} Proposition 1.4.2) states that any digraph that has no vertex of in-degree 0 must have a directed cycle. 
Now if $r \in \R'$ then $r$ has a catalyst in ${\rm cl}_{\R'}(F)$ and so this catalyst is either the product of some reaction in $\R'$ or it is in $F$. However, the latter possibility  is ruled out
by the stated assumption concerning $\R'$. Thus each vertex of $G'(\R')$ has positive in-degree and so this digraph has a directed cycle.

{\em Part (ii):}  Suppose $\R'$ is the smallest RAF for $(\Q,F)$. From  part (i) $\R'$ contains a directed cycle of some length $k$,  so $|R'| \geq k$.  
Moreover, since $\R' \subseteq s(\R)$, 
$k$ is at least the size of the smallest directed cycle in $G'(s(\R))$, as claimed.  Moreover, since $s(\R)$ can be computed in polynomial time (by the RAF algorithm from Section~\ref{rafalg}), and thus $G'(s(\R))$ can be also, one can find the shortest directed cycle in this graph by an application of the Floyd--Warshall algorithm, or via Dijkstra's algorithm (see, for example,  \cite{ban}).

\section{Minimal RAFs in the binary polymer model}

To investigate the issue of the smallest RAFs empirically, we used the {\em binary polymer model} to collect statistics on the sizes of RAF and irrRAF sets.  This model has all binary sequences of length at most $n$ as its molecules, and the reactions consist of ligation reactions (joining two sequences to form a longer sequence), together with the reversal of this operation (cleavage reactions, in which a sequence is split into two subsequences).  Examples of ligation and cleavage reactions are $0101+001 \rightarrow 0101001$ and $11110 \rightarrow 111+10$, respectively.

In this model, a ligation reaction and its associated cleavage reaction are often regarded as the same (reversible) `cleavage-ligation' reaction.  We let $\R=\R_n$ denote this set of cleavage-ligation reactions, and for a subset $\R'$ of $\R_n$, the set $\pi(\R')$ will be taken to be the set of of products of the cleavage and ligation reactions associated with $\R'$ (and so the closure of $F$ relative to $\R'$  is the closure of $F$ relative to the union of the associated cleavage and ligation reactions).   

In the simplest form of this model, each molecule $x$ catalyses any given cleavage-ligation reaction $r$ independently with probability $p=p_n$, which depends on $n$. The food set $F$ is usually chosen to be all binary sequences of length at most $t$ for a small value of $t$ (typically, $t=2$, in which case $|F|=6$).  

In previous work, we already studied how the probability of RAF sets existing in this model scales with the value of $n$ (the maximum length of molecules). Here, we simply chose one value ($n=10$) and computed the sizes of RAF sets for various values of $p$ (the probability that a given molecule catalyses a given cleavage-ligation reaction) or, equivalently, the level of catalysis $f=p|\R|$ (the average number of reactions catalysed per molecule).

\begin{figure}[htb]
\centering
\includegraphics[scale=.5]{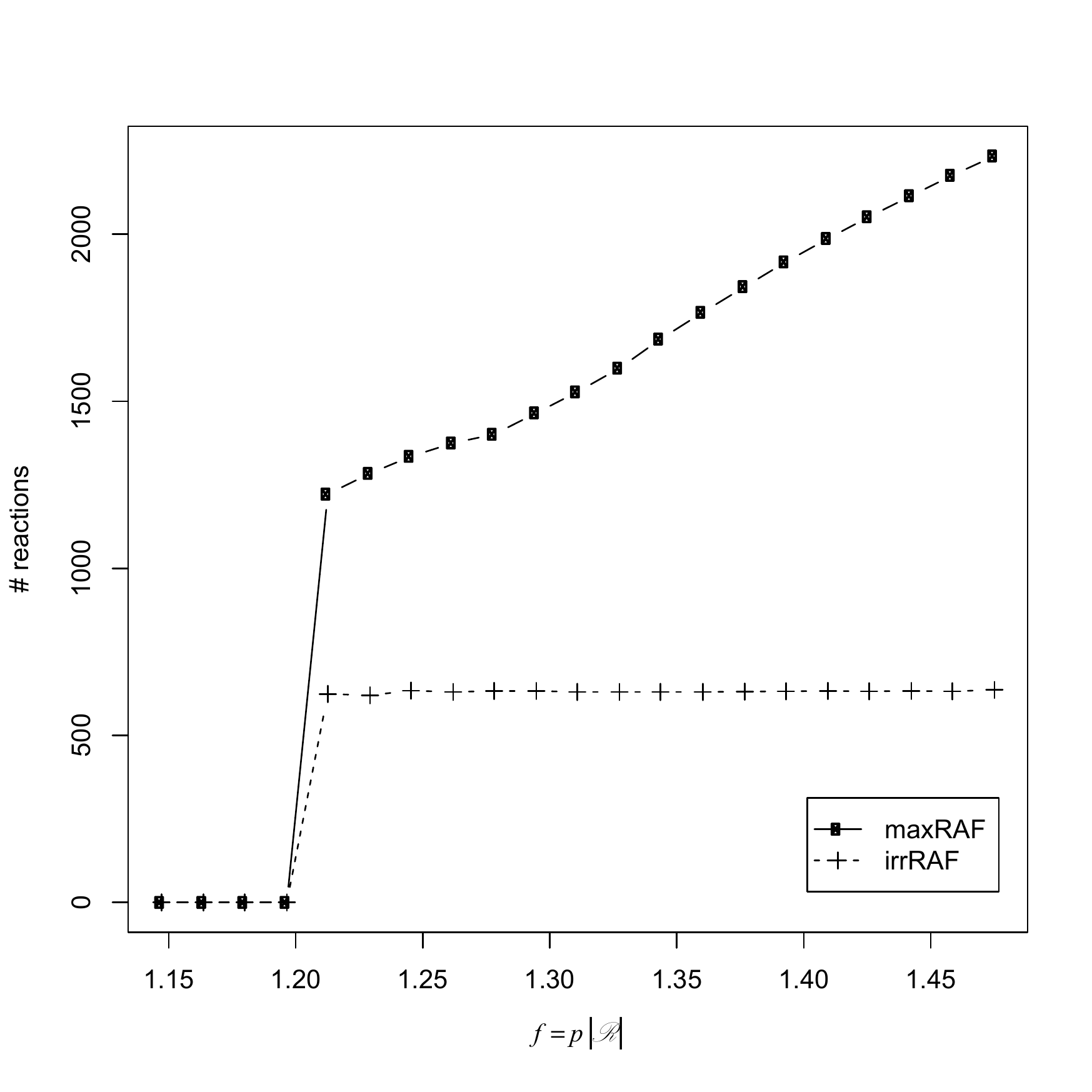}
\caption{The average sizes of RAF and irrRAF sets for increasing levels of catalysis for $n=10$ in the binary polymer model.}
\label{fig:RAFsize}
\end{figure}

Fig. \ref{fig:RAFsize} shows the average sizes of RAF sets (black squares) and irrRAF sets (crosses) for increasing levels of catalysis. These data points are averages over 1000 instances of the model for each value of $p$. When the level of catalysis is too low ($f<1.20$), no RAF sets are found at all, i.e., their sizes are equal to zero. However, at a level of catalysis just above $f=1.20$, the first RAF sets are starting to show up. Initially, they are found in only 6 out of 1000 model instances, but with increasing levels of catalysis $f$, they become more and more frequent, and their sizes seem to increase linearly with $f$. In contrast, the average size of irrRAFs remains constant (for each non-empty RAF set, one (arbitrary) irrRAF set was generated) as the rate of catalysis increases across this narrow interval. 

An interesting feature of Fig. \ref{fig:RAFsize}  is that the sizes of the RAF sets when they first start appearing (around $f=1.20$) are already quite large: 1222 reactions on average in the six RAF sets and 624 reactions in the corresponding six irrRAF sets (with $|\R|=16388$ for the full reaction set). So, it seems there are no ``small'' RAFs when they are only just starting to appear. This observation is formalised in the following theorem, which shows that at the catalysis levels at which RAFs have a moderate probability of occurring, 
 the smallest RAFs  have a size that grows exponentially with  $n$.

\begin{theorem} [Threshold catalysis RAFs  have exponential size in $n$]
\label{nosmall}
\mbox{ }

\noindent Consider the binary polymer model $\Q_n$ for sequences up to length $n$. 
Select any fixed value $v<1$ and then select the catalysis probability $p=p_n$ so that $$Pr(\exists  \mbox{  RAF  }  \mbox{ for }  \Q_n) = v.$$
Then, for any constant $c<\frac{1}{3}$:
$$Pr(\exists  \mbox{  RAF }  \R' \mbox{ for } \Q_n:   |\R'| \leq 2^{cn}) \rightarrow 0,$$
 as $n \rightarrow \infty$.

\end{theorem}
{\em Proof:}  For any subset $\R'$ of $\R_n$ with  $s= |F \cup \pi(\R')|$  we have $|{\rm cl}_{\R'}(F)| \leq s$, and so  
the probability that an arbitrary reaction $r \in \R'$ is catalysed by
at least one element of ${\rm cl}_{\R'}(F)$  is at most $1-(1-p_n)^s$.  Consequently, if, in addition, $\R'$ has size $k$, 
the probability that $\R'$ is reflexively autocatalytic is  at most
$(1-(1-p_n)^s)^k.$ 
Now, we can provide a further upper bound on this last probability by an expression that involves just $k$ (and not $s$) by observing that:
\begin{equation}
\label{list}
(1-(1-p_n)^s)^k \leq (sp_n)^k \leq [(3k+|F|)p_n]^k,
\end{equation} 
by noting that $s \leq 3k+|F|$, since each reaction in $\R'$ is associated with at most three distinct molecules.
  
In summary, the probability that any subset $\R'$ of $\R_n$ of size $k$ is reflexively autocatalytic is, at most:
\begin{equation}
\label{upper_bound}
[(3k+|F|)p_n]^k.
\end{equation}
Let $S_{n,k}$ be the number of subsets of $\R_n$ of size $k$ that are $F$-generated. Boole's inequality, combined with the upper bound (\ref{upper_bound}), implies that 
 the probability that $\Q_n$ has an RAF of size $k$ is bounded above by:
$S_{n,k}\cdot [(3k+|F|)p_n]^k$.
Thus:
$$Pr(\exists  \mbox{  RAF }  \R'  \mbox{ for } \Q_n:   |\R'| \leq m) \leq \sum_{k=1}^{m} S_{n,k} \cdot  [(3k+|F|)p_n]^k.$$
Now, the value of $p_n$ for which $Pr(\exists  \mbox{ RAF  } \mbox{ for }  \Q_n) = v$ is bounded above by $\lambda_v n /|\R_n|$ for some value $\lambda_v$ dependent only on $v$ (by \cite{mos} [Theorem 4.1], and   \cite{hor} [Proposition 8.1]).
Thus:
\begin{equation}
\label{peq}
Pr(\exists  \mbox{  RAF }  \R'\mbox{ for } \Q_n:   |\R'| \leq m) \leq \sum_{k=1}^{m} S_{n,k} \cdot  [(3k+|F|)\lambda_v n/|\R_n|]^k.
\end{equation}

Now, by Lemma \ref{lemequiv2}, any set of reactions  is $F$-generated if and only if the reactions can be linearly ordered so that every reaction in the sequence has its reactants provided either from $F$ or from the products of earlier reactions in the sequence (or both).

Therefore, $S_{n,k}$ is bounded above by the collection of ordered sequences $r_1, r_2, \ldots, r_k$ where, for all $j: 0 \leq j < k$:
\begin{itemize}
\item[(*)]  $r_{j+1}$ is a cleavage or ligation reaction involving one or two (respectively) molecules of $X_j:=F \cup \pi(\{r_1, \ldots, r_j\})$
(taking $X_0 = F$).
\end{itemize}
Now, each reaction in the sequence $r_1, r_2, \ldots, r_k$ creates, at most, two new molecules, and so $|X_{j+1}| \leq |X_j|+2$ for all $j$.  
Since $X_0=F$, we have for all $0 \leq j \leq k-1$:
\begin{equation}
\label{ups}
|X_j|\leq |F|+2j,
\end{equation}
Now, given $r_1, \ldots, r_j$ (where $j<k$), the number of possible choices for $r_{j+1}$ to satisfy condition (*) above is,  at most:
 $$|X_j|^2 + n\cdot |X_j|,$$
since the first term in this sum is an upper bound on the number of possible ligation reactions, while the second term is an upper bound on the number of cleavage reactions.
Combining this with (\ref{ups}) gives the following upper bound on the number of sequences 
$r_1, r_2, \ldots, r_k$ satisfying (*).
$$\prod_{j=0}^{k-1} \left[(|F|+2j)^2+n(|F|+2j) \right] \leq \left[(|F|+2k)(n+|F|+2k)\right]^k \leq (n+|F|+2k)^{2k},$$
and so 
$$S_{n,k} \leq  (n+|F|+2k)^{2k}.$$
Applying this inequality to (\ref{peq}), with the asymptotic equivalence $|\R_n| \sim n 2^{n+1}$, gives:
\begin{equation}
\label{sumeq}
Pr(\exists  \mbox{  RAF }  \R' \mbox{ for } \Q_n:   |\R'| \leq m) \leq \sum_{k=1}^{m}  [(3k+|F|)\lambda_v (n+|F|+2k)^2/2^{n+1}]^k.
\end{equation}
Notice that we can provide an upper bound  for the term on the right by the expression:
 $$\sum_{k=1}^{\infty}  [(3m+|F|)\lambda_v (n+|F|+2m)^2/2^{n+1}]^k = \theta/(1-\theta),$$
 where $\theta = [(3m+|F|)\lambda_v (n+|F|+2m)^2/2^{n+1}]$.  It follows that 
 if $m\leq 2^{cn}$  for $c< \frac{1}{3}$, then $\theta$ (and thereby $\theta/(1-\theta)$)  converges to zero as $n \rightarrow \infty$,
 and therefore so too does the expression for the probability in (\ref{sumeq}). 
This completes the proof.

{\bf Comments}
\begin{itemize}
\item
This result is interesting in the light of Theorem 11 of \cite{bol}, as the probability that the length of a first cycle is $k$ when a first cycle appears in a random digraph is 
$1/k(k+1) +o(1)$, and so short cycles have considerable probability in that model.

By contrast, when the first RAFs appear, there are no small ones, since any RAF requires the simultaneous satisfying of two properties:
it must be reflexively autocatalytic and
also $F$-generated; the former property is equivalent to the existence of a directed cycle in the catalysis graph (at least in the case $p(x,r)=0$ for $x\in F$);  
while there might be a small cycle, it is unlikely to be $F$-generated.

\item Theorem \ref{nosmall} provides an interesting complement to the earlier Theorem~\ref{npthm}, which showed that there is, in general, no efficient way to determine the size of the smallest RAF in a CRS.  Thus, it could be difficult to exclude the possibility a small RAF in  the binary polymer model for large values of $n$, 
by searching for the smallest irrRAFs. However,  Theorem~\ref{nosmall}  provides a theoretical guarantee that, with high probability, there will be no small RAFs when they first appear within this model.

\item  The final inequality in the  proof of Theorem \ref{nosmall} allows us to place explicit bounds on the likely minimal size of RAFs for finite values of $n$.  For example, for
$n=40$, the probability that there exists an RAF of size 1000 when the existence of an RAFs has a probability of 0.5 is less that 0.01 (taking $|F|=6$ and the conservative value for $\lambda_v$ of $1.7$ from Theorem 4.1(ii) of \cite{mos}).

\item 
 It is easy to show that when the rate of catalysis becomes sufficiently large, we will expect to find small RAFs in the binary polymer model. Thus the initially largely flat line
 for irrRAF sizes in Fig.~\ref{fig:RAFsize} must eventually decrease to small values (in the limit of size 1) as the rate of catalysis continues to increase.   Moreover, 
 small catalytic reaction systems (of size 16) that form RAFs (and which contain even smaller RAFs) have  recently been discovered in real RNA replicator systems \cite{vai}. That such small sets form RAFs can be partly explained by the high catalysis rate \cite{hor4}.

\end{itemize}

\subsection{Distribution of irrRAF sizes}

With Theorem \ref{npthm} above, we proved that finding the smallest (irr)RAF set is a hard problem, so we cannot hope to have a polynomial time algorithm to do this. However, it is still possible to get an idea of the distribution of the sizes of the irrRAF sets that exist inside an RAF set. This can be done as follows. In \cite{hor}, we described a polynomial time algorithm for finding one possible irrRAF in a given RAF $\R'$ by removing one reaction $r_i$ from $\R'$ and applying the RAF algorithm to the set $\R'-\{r_i\}$. If this results in an empty set ($s(\R'-\{r_i\})=\emptyset$), then reaction $r_i$ is essential and needs to remain in $\R'$. Otherwise, replace $\R'$ by the non-empty subRAF $s(\R'-\{r_i\})$. Now repeat this procedure with every next reaction $r_i$ in $\R'$ until all reactions have been considered. The result of this is an irrRAF of $\R'$. This algorithm was used to generate the data on irrRAF sizes in Fig. \ref{fig:RAFsize}.

\begin{figure}[htb]
\centering
\includegraphics[scale=.33]{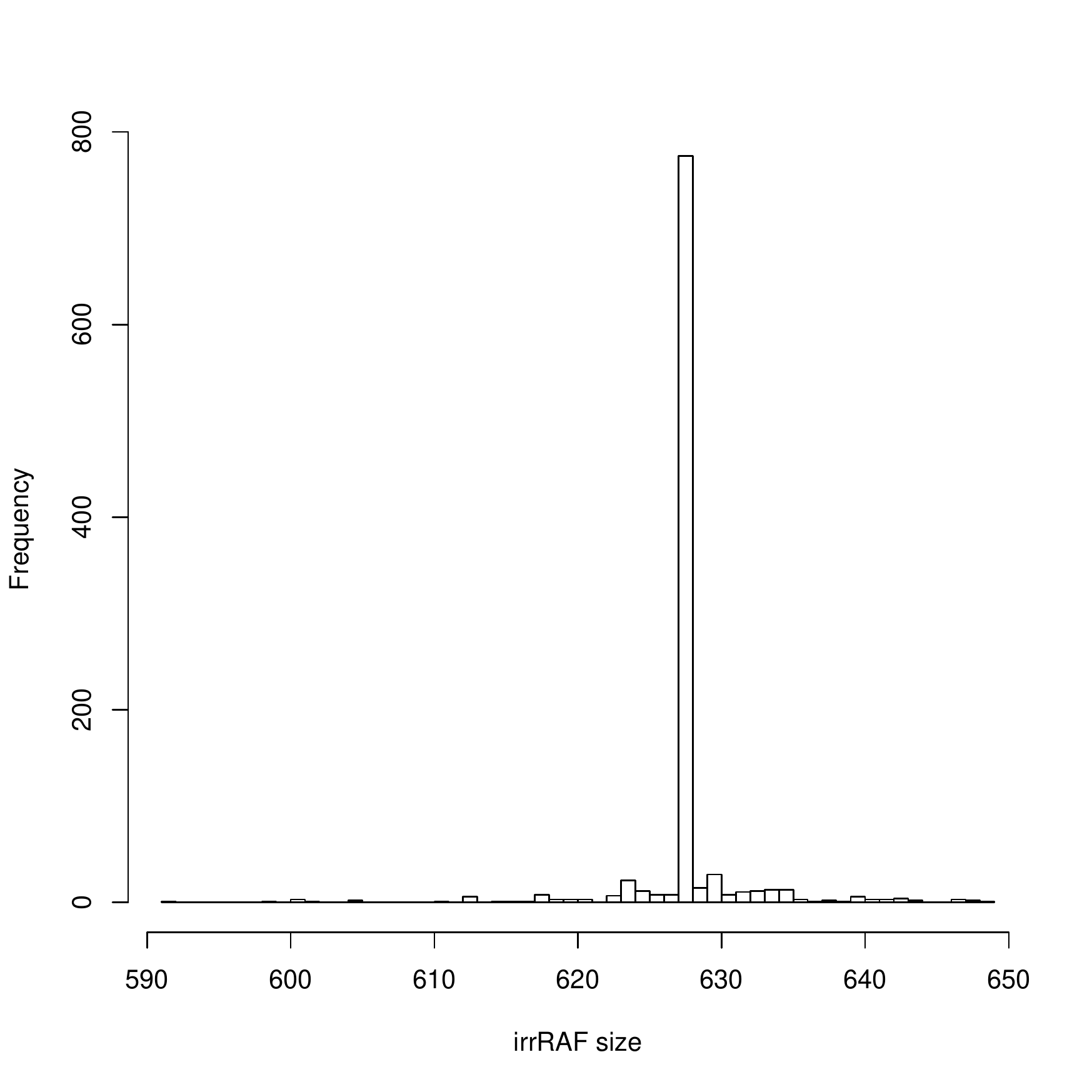}
\includegraphics[scale=.33]{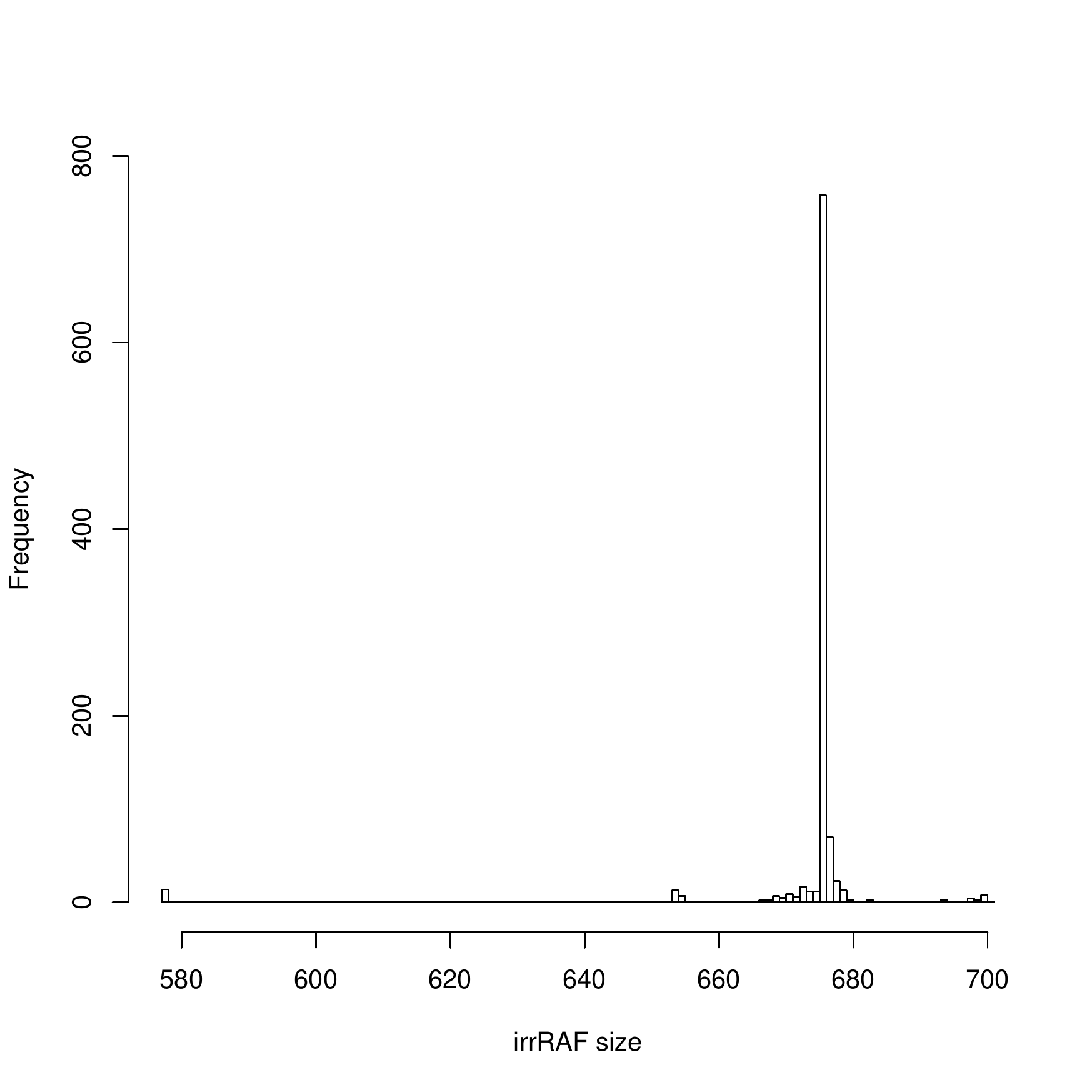}
\caption{Histograms of the sizes of 1000 irrRAFs in two RAF sets when they first start to appear in the binary polymer model.}
\label{fig:irrRAF}
\end{figure}

Note that the particular irrRAF in $\R'$ that is found by this algorithm depends on the order in which the reactions $r_i \in \R'$ are considered for possible removal. So, by repeating the above algorithm a number of times and  randomly re-ordering the reactions in $\R'$ each time, we can generate a sample of irrRAFs of $\R'$. Fig. \ref{fig:irrRAF} shows two histograms of the sizes of 1000 irrRAFs generated this way from two of the RAF sets that were found at a level of catalysis of about $f=1.20$, i.e., when RAF sets are just starting to show up.

In both cases, the sample is dominated by one particular irrRAF size, with the rest being relatively close in size, although the histogram on the right shows a case where the smallest irrRAF is about 100 reactions smaller than the dominant one. Since this is only a random sample, there is no guarantee that this is indeed the {\it smallest} irrRAF. However, the fact that even the smallest irrRAF in these samples is still rather large (close to 600 reactions) is probably a good indication that, indeed, there are no small RAFs when they just start appearing.

\section{Concluding comments}

RAF theory provides a way to address one aspect of the complex question, how did life arise? The existence of RAFs does not represent a sufficient condition, but it would seem to be a necessary one.  Moreover, the approach is sufficiently general that it can be applied to other emergence phenomena both inside chemistry and in quite disparate fields (for an application to a `toy' problem in economics, see \cite{hor_ext}).  RAFs are based on two key ideas -- every molecule must be able to be built up from the available set of `food' molecules by reactions from the set, and each reaction must `eventually' be catalysed. Here `eventually' refers to the fact that some reactions may need to proceed uncatalysed (at a lower rate) in order to get the system going, but eventually, all reactions are catalysed. A stronger requirement would be that all reactions must be catalysed by the available molecules as the system  develops (from the food molecules or products of reactions that have already occurred). This notion of a `constructively autocatalytic $F$-generated' (CAF) set from  \cite{mos} seems an unnecessarily strong condition (since reactions can generally proceed, at a lower rate, without catalysis) and the mathematical properties of CAFs (and the probability that they form) are quite different from RAFs \cite{mos}.  A weaker requirement is that only some reactions need to be catalysed -- this fits perfectly easily within the current RAF framework, as we may simply formally allow a food molecule  to act as a putative catalyst for those reactions.

Another weakening of the RAF concept is to consider a closed chemical reaction system, which,  once established, will continue to be self-maintaining. This underlies the notion of an `organisation' in chemical organisation theory.  The property of RAFs of being $F$-generated was shown in \cite{con} to imply the property of being an organisation; we have shown here that the converse need not hold -- in other words, an organisation may not be able to be built starting just with the food set, without the presence of some other reactant to get it started.
This property of an organisation has a superficial similarity to the property that a RAF can allow one or more some reactions to proceed uncatalysed until the catalyst is formed.  However, there is an important difference, since an uncatalysed reaction can proceed (at a lower rate),  while this a reaction that lacks one of its reactants cannot take place.

The focus of this paper has been on small RAFs, as these are, in some sense, the `simplest' systems that could be of interest in origin-of-life studies. It is of interest to know whether within some CRS that harbours an RAF, there is a very small one present, or instead whether all subRAFs are quite large. 
The smallest RAFs are irreducible, though not all irreducible RAFs are of the smallest size.  In contrast to the maximal RAFs, where there is a unique object (maxRAF)  that can be  constructed in polynomial time (by the RAF algorithm), there may be exponentially many irreducible RAFs, and finding a smallest RAF is, in general, NP-hard.  Nevertheless, we can find irrRAFs in polynomial time, and we can describe computable lower bounds on the size of irrRAFs and also determine if a given (small) collection comprises all the irrRAFs.  

It  is also of interest to consider the size and distribution of RAFs in simple settings such at the binary polymer model, where simulations suggest that when RAFs first appear, small irrRAFs are unlikely, a result that has been verified formally in Theorem~\ref{nosmall}.  However, as the level of catalysis increases, one is guaranteed to eventually find small irrRAFs. 

An interesting problem for future work would be to develop better bounds and approximations for  the minimal size of a RAF within a catalytic reaction system. For example, is it possible to obtain a bound for the size the smallest RAF that is within some constant factor of optimal?  It would also be of interest to investigate an extension of RAFs that allow some molecules to not only catalyse some reactions, but also to inhibit other reactions; in this case determining whether an analogue of an RAF exists within an arbitrary  CRS has been shown to NP-hard \cite{mos}, but in certain cases
the RAF algorithm can be adapted to solve this problem \cite{hor_ext}.

\section{Acknowledgments}
We thank the {\em Allan Wilson Centre for Molecular Ecology and Evolution} for helping fund this work.

\section{Appendix}

\subsection{ Proof of Proposition~\ref{proprop}}

To establish $(ii) \Rightarrow (i)$, suppose that $s(\R - \R') \neq \emptyset$ and that $\R' \cup s(\R - \R')$ is an RAF for $(\Q,F)$. $s(\R-\R')$ is an RAF for $(\Q',F)$, where $\Q'=\{X,\R - \R',C - C'\}$ with $C':=\{(x,r)| (x,r) \in C, r \in \R '\}$, and so is certainly an RAF for $(\Q ,F)$. Furthermore $\R' \cap s(\R - \R') = \emptyset$, since $s(\R - \R') \subseteq \R - \R'$. Hence $\R'$ is a co-RAF for $(\Q,F)$.

To establish $(i) \Rightarrow (ii)$ suppose that $\R'$ is a co-RAF for $(\Q,F)$. Then there exists an RAF $\R_1$ for $(\Q,F)$, such that $\R' \cap \R_1 = \emptyset$ and $\R' \cup \R_1$ is an RAF for $(\Q,F)$. Consider $s(\R-\R')$. Since $\R_1$ is an RAF for $(\Q,F)$ and is a subset of  $s(\R-\R')$, we must have $s(\R-\R') \neq \emptyset$.  It remains to show that
$\R' \cup s(\R-\R')$ is an RAF for $(\Q,F)$. Suppose that $r \in \R' \cup s(\R-\R')$.  Then either $r\in \R'$, in which case all the reactants of $r$ and at least one catalyst are contained in ${\rm cl}_{\R' \cup \R_1}(F)$ (since $\R' \cup \R_1$ is an RAF for $(\Q,F)$), while if $r \in s(\R-\R')$ then all the reactants of $r$ and at least one catalyst is contained in ${\rm cl}_{s(\R-\R')}(F)$ (since $s(\R-\R')$ is an RAF for $(\Q,F)$).  Now,  $\R' \cup \R_1$ and $s(\R-\R')$ are both subsets of $\R' \cup s(\R-\R')$, and so ${\rm cl}_{\R' \cup \R_1}(F)$ and ${\rm cl}_{s(\R-\R')}(F)$
are both subsets of ${\rm cl}_{\R' \cup s(\R-\R')}(F)$. Consequently, every reaction in $\R' \cup s(\R - \R')$ has all its reactants and at least one catalyst in ${\rm cl}_{\R' \cup s(\R-\R')}(F)$,
which implies that $\R' \cup s(\R - \R')$ is an RAF for $(\Q,F)$.

To establish $(iii) \Rightarrow (i)$, note that $\R_A$ is an RAF for $(\Q,F)$ such that $\R' \cap \R_A = \emptyset$ and $\R' \cup \R_A = \R_B$ which is an RAF for $(\Q,F)$. 
Therefore, $\R'$ is a co-RAF for $(\Q,F)$.

To establish $(i) \Rightarrow (iii)$, suppose that $\R'$ is a co-RAF for $(\Q,F)$. Then there exists an RAF $\R_1$ for $(\Q,F)$, such that $\R'\cap \R_1 = \emptyset$ and $\R' \cup \R_1$ is an RAF for $(\Q,F)$. Trivially, $\R' = (\R' \cup \R_1) - \R_1$ and clearly $\R_1\subset \R'  \cup \R_1$, since $\R'$ is non-empty by  the definition of a co-RAF, so take $\R_A = \R_1$ and $\R_B = \R' \cup \R_1$.

To establish $(i) \Rightarrow (iv)$, suppose that $\R'$ is a co-RAF for $(\Q,F)$. Then there exists an RAF $\R_1$ for $(\Q,F)$ such that $\R_1 \cap \R' = \emptyset$ and $\R_1 \cup \R'$ is an RAF for $(\Q,F)$. It suffices to show that $\R'$ is an RAF for $(\Q,F')$, where $F' = F \cup \pi(\R_1)$. First we prove that $\R'$ is generated from $F'$: i.e.  $\rho(\R') \subseteq {\rm cl}_{\R'}(F')$. Let $m = |\R_1|, n = |\R'|$. $\R_1 \cup \R'$ is $F$-generated, so there exists an ordering $O_u = u_1, \dots, u_{m+n}$ of its reactions $u_i$ satisfying part (iv) of Lemma~\ref{lemequiv2} for the food set $F$. We herein refer to an ordering satisfying part (iv) of Lemma~\ref{lemequiv2} for some food set $F$ as a {\em proper ordering relative to F}. $\R_1$ is $F$-generated so there exists a proper ordering relative to $F$,  $O_1 = r_1, \dots, r_m$, of its reactions $r_i$. Define $O' = r'_1, \dots, r'_n$ to be the ordering of the reactions of $\R'$ obtained by deleting from $O_u$ every reaction that also appears in $O_1$, preserving the order of the remaining reactions. We claim that the concatenation $O_1, O'$, a reordering of $O_u$, is a proper ordering relative to $F$. Consider any reaction $r' \in \R'$ and a reactant $x \in \rho(r')$. $\R_1 \cup \R'$ is $F$-generated and $\R' \subset \R_1 \cup \R'$, so by part (ii) of Lemma~\ref{lemequiv2} at least one of the following holds: (i) $x \in F$, (ii) $x \in \pi(r)$ for some $r \in \R_1$, or (iii) $x \in \pi(r'')$ for some $r'' \in \R'$. If (i) alone is true, $r'$ trivially does not prevent the reordering from being a proper ordering relative to $F$. If (ii) alone is true, every $r \in \R_1$ precedes $r'$ in $O_1, O'$, so $r'$ certainly does not prevent the reordering from being a proper ordering relative to $F$. If (iii) alone is true, $r''$ must precede $r'$ in $O_u$ and the order of the reactions of $\R'$ in $O_u$ is preserved in $O_1, O'$, so $r''$ precedes $r'$ in $O_1, O'$. If more than one of (i)-(iii) are true, then since $O_u$ is a proper ordering, at least one of the conclusions will hold, which is sufficient. Therefore our claim that $O_1, O'$ is a proper ordering relative to $F$ is justified. It follows that $\rho(r'_1) \subseteq F \cup \pi(\R_1)$, and for each $i \in \{2, \dots, n \}$, $\rho(r'_i) \subseteq F \cup \pi(\R_1) \cup \pi( \{ r'_1, \dots, r'_{i-1} \})$, so moreover $O'$ alone is a proper ordering relative to $F'$. Then, by the implication $(iv) \Rightarrow (i)$ in Lemma 3.1, $\R'$ is generated from $F'$. It remains to show that $\R'$ is reflexively autocatalytic. Since $\R'$ and $\R_1 \cup \R'$ are $F$-generated, we can apply part (ii) of Lemma~\ref{lemequiv2} to ${\rm cl}_{\R'}(F')$ and ${\rm cl}_{\R_1 \cup \R'}(F)$ to deduce that they are equal. Now, since $\R_1 \cup \R'$ is reflexively autocatalytic  then certainly $\R'$ is reflexively autocatalytic (by definition).

To establish $(iv) \Rightarrow (i)$, it suffices to show that $\R_1 \cup \R'$ is an RAF for $(\Q,F)$, since we already have  that $\R_1$ is an RAF for $(\Q,F)$ and $\R_1 \cap \R' = \emptyset$. First we prove that $\R_1 \cup \R'$ is $F$-generated. $\R_1$ is $F$-generated, so there exists a proper ordering relative to $F$ of its reactions $ r_1, \dots , r_m $. Similarly for $\R'$ there exists a proper ordering relative to $F \cup \pi(\R_1)$ of its reactions $r'_1, \dots, r'_n$. Hence the concatenation $ r_1, \dots, r_m, r'_1, \dots, r'_n$ is a proper ordering relative to $F$ of the reactions in $\R_1 \cup \R'$, so $\R_1 \cup \R'$ is $F$-generated. It remains to show that $\R_1 \cup \R'$ is reflexively autocatalytic. Since $\R_1, \R'$ and $\R_1 \cup \R'$ are each $F$-generated, we can apply part (ii) of Lemma~\ref{lemequiv2} to each of ${\rm cl}_{\R_1}(F)$, ${\rm cl}_{\R'}(F')$ and ${\rm cl}_{\R_1 \cup \R'}(F)$ to deduce that ${\rm cl}_{\R_1}(F) \subseteq {\rm cl}_{\R'}(F') = {\rm cl}_{\R_1 \cup \R'}(F)$. Now since $\R_1$ and $\R'$ are reflexively autocatalytic then certainly $\R_1 \cup \R'$ is reflexively autocatalytic. This completes the proof.

 \subsection{Proof of Theorem~\ref{npthm}}

{\em Proof:}  MIN-RAF is clearly in the complexity class NP, since one can  verify in polynomial time if a given subset of $\R$ has size, at most, $k$ and forms an RAF.
We will reduce the graph theory problem VERTEX COVER to MIN-RAF. Recall that for a graph $G=(V,E)$, a {\em vertex cover} of $G$ is a subset
$V'$ of $V$ with the property that each edge of $G$ is incident with at least one vertex in $V'$; 
VERTEX COVER has as its instance a graph $G=(V,E)$ and 
an integer $K$ and we ask whether or not $G$ has a vertex cover of size, at most, $K$.
This is a well-known NP-complete problem \cite{gar} (indeed, one of Karp's original 21 NP-complete problems). 
Given an instance $(G=(V,E), K)$ of VERTEX COVER, we show how to construct an instance $(X_G, \R_G, C_G, F_G, k)$, of MIN-RAF for which the answers to the two decision problems are identical.

We first construct $F_G$ and $X_G$.   For each  $v\in V$, let $a_v, b_v$ be two distinct elements of $F_G$ and let $x_v$ be an element of $X_G-F_G$.
Order $E$ as $e^1, \ldots, e^{|E|}$ and for each $j=1, \ldots, |E|$, let $d_j$ be a distinct element of $F$ and $y_j$ an element of $X_G-F_G$.
Let $d_0$ be another distinct element of $F_G$.   Thus $F_G$ consists of the $2|V|+|E|+1$ elements:
$$F_G:= \{d_j: 0 \leq j \leq |E|\} \cup \{a_v, b_v: v \in V\}$$
and $X_G-F_G$ consists of $|V|+ |E|$ elements:
$$X_G-F_G:= \{x_v: v \in V\} \cup \{ y_j: 1\leq j \leq |E|\}.$$
 \begin{figure*}[h] 
 \center
\resizebox{10cm}{!}{
\includegraphics{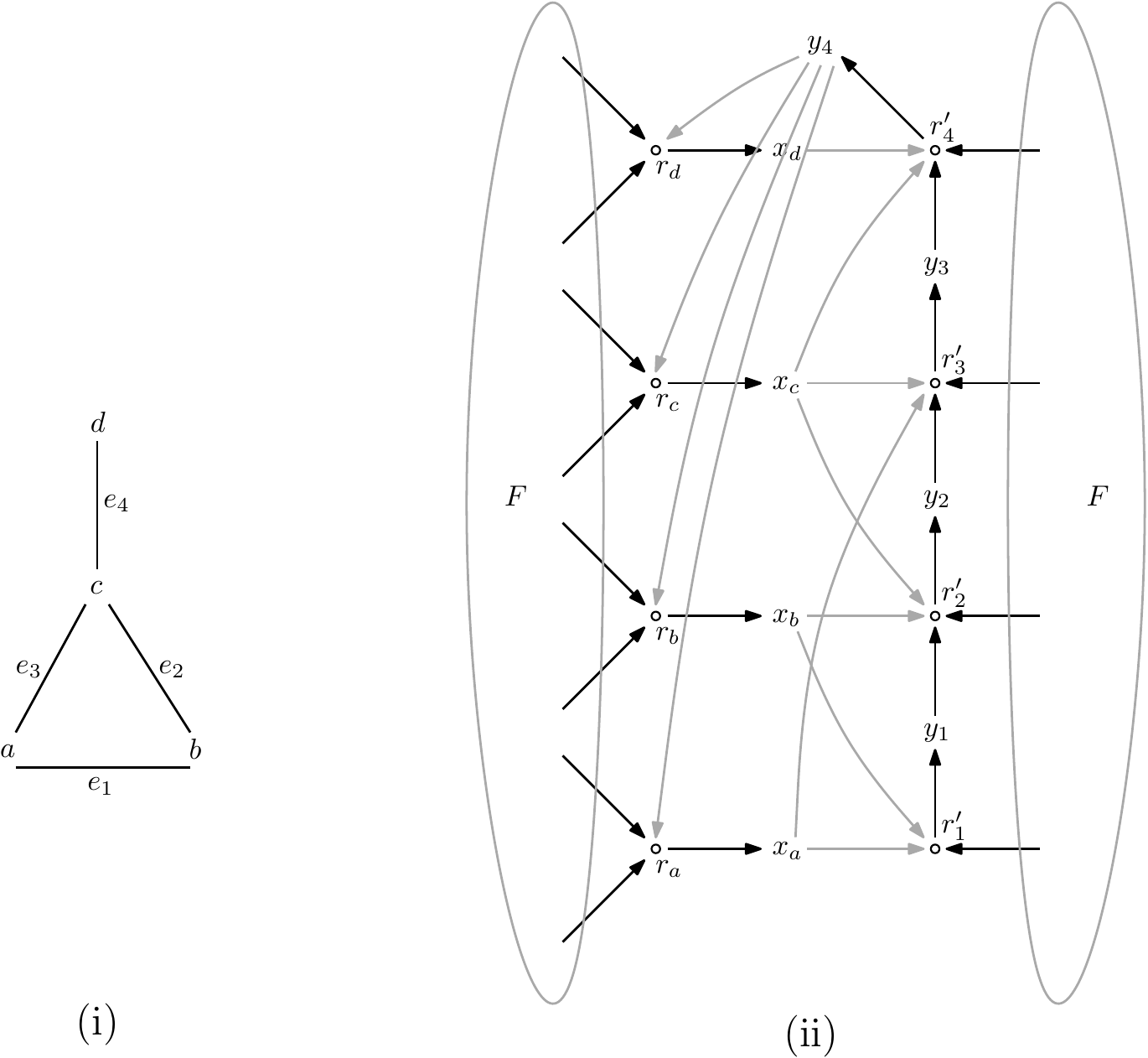}
}
\caption{(i) A graph $G$ and (ii) the associated CRS $\Q_G$, consisting of 8 reactions that form an RAF, and with the 
 super-catalyst ($y_4$) at the top.  The two smallest sub-RAFs of this system are formed by adding either  $r_a$ and $r_c$ or
$r_b$ and $r_c$  to the four reactions $r'_1, \ldots, r'_4$, and these two choice correspond to the two smallest vertex covers of $G$, namely $\{a,c\}$ and $\{b,c\}$.
}
\label{figure0}
\end{figure*}
For each $v \in V$, define a reaction: $$r_v: a_v + b_v \rightarrow x_v.$$
For  each $1<j \leq |E|$,  define the reaction:
$$r'_j:  y_{j-1}+d_j \rightarrow y_j,$$
and for $j=1$, let:
$$r'_1: d_0+d_1 \rightarrow y_1.$$ 
For any subset $U$ of $V$  let:
$$\R_U = \{r_v: v \in U\}, \mbox{ and let } $$
$$\R_V := \{r_v: v \in V\} \mbox{ and } \R_E := \{r'_j: 1\leq j \leq |E|\},$$
and set $$\R_G = \R_V \cup \R_E.$$

Thus, we have specified  $X_G, F_G$ and $\R_G$ and it remains to define the catalysis ($C_G$) assignment, which is as follows:
\begin{itemize}
 \item
If $e^j = (u^j, v^j)$ (where $u^j, v^j \in V$)  then $r'_j$ is catalysed by both $x_{u^j}$ and $x_{v^j}$ (but by no other molecules).  
\item
In addition, each reaction
$r_v: v\in V$ is catalysed by $y_{|E|}$ and by no other molecule - we call the molecule  $y_{|E|}$ the {\em super-catalyst}.
\end{itemize}

An example of this construction is illustrated in Fig.~\ref{figure0}.
We have now fully specified the catalysation and thereby the pair $(\Q_G, F_G)$ constructed from $G$ ($\Q_G = (X_G, \R_G, C_G)$).

\noindent {\bf Claims:}  

\begin{itemize}
\item $\R_G$ is an RAF for $(\Q_G, F_G)$.
\item A  subset $\R'$ of $\R_G$ is an RAF for $(\Q_G, F_G)$ if and only if $\R' = \R_{V'} \cup \R_E$ for a vertex cover $V'$ of $G$.
\item  The vertex covers of $G$ of size $K$ are in one-to-one correspondence with the sub-RAFs of $\R_G$ of size $K+|E|$. 
\end{itemize}

The first claim is readily verified. 

To establish the second claim, suppose that  $V'$ is a vertex cover of $G$. Then every reaction in $\R_E$ is catalysed by the product of least one reaction in
$\R_{V'}$. Moreover, the product of $r'_{|E|}$ catalyses all the remaining reactions.  Thus, $\R'$ is reflexively autocatalytic, and it is 
also clear that $\R'$ is $F$-generated; thus  $\R'$ is an RAF and it has $K+|E|$ reactions.
Conversely, suppose that $\R''$ is an RAF for $(\Q_G, F_G)$ of size at most $K+|E|$.
If $r'_{E}$ is not in $\R''$ then the super-catalyst is not produced by any reaction in  $\R''$ so none of the reactions in $\R_V$ is catalysed; moreover, because the products from these last reactions provide the only catalysts for $\R_E$ it follows that $\R'' = \emptyset$. Thus, since $\R''$ is non-empty (being an RAF),  $r'_{|E|}$ must be an element of $\R''$, and in order to construct the reactants of $r_{|E|}$,  all the reactions $\R_E$ must form a subset
of $\R''$. In order for all these reactions to be catalysed, at least one of the reactions $r_{u^j}$ and $r_{v^j}$ must lie in $\R''$
for each $1 \leq j \leq |E|$. Thus $\{v: r_v \in \R''\}$ is a vertex cover of $G$ and it has size, at most, $(K+|E|)-|E| = K$ as claimed. 
This establishes the required reduction, and thereby completes the proof of the second claim.

The third  claim follows by the noting that the association $V' \mapsto \R_{V'} \cup \R_E$ maps vertex covers of $G$  of  size $K$  onto sub-RAFs of $\R_G$ of size $K+|E|$
(by the previous claim) and two different vertex covers are mapped to distinct sub-RAFs.  This completes the proof.

Part (i) of Theorem~\ref{npthm} now follows from the first two claims, while Part (ii) of Theorem~\ref{npthm}  follows from the third claim, combined with the \#P-completeness of counting vertex covers of a graph and minimum vertex covers of a graph (see \cite{vad}). 

\bigskip

{\bf Remark:}  We have ensured in the proof above that each reaction has just two reactants, in line with the binary polymer model. 
However, the attentive reader will notice that $F$ may have to be quite large.
Nevertheless, it is quite straightforward to modify this example so that $F$ is kept small (e.g. of size 6), and to implement the construction within the constraints of the binary polymer cleavage--ligation model.

\end{document}